\documentclass[10pt,english,aps,prb,preprint,showpacs]{revtex4}
\usepackage{amsmath}
\usepackage{graphicx}
\usepackage{color}

\usepackage{bm}
\usepackage{babel}
\makeatother
\begin{document}

\title{Identification of long-lived clusters and their link to slow dynamics in a model glass former}

\author{Alex Malins}
\affiliation{Bristol Centre for Complexity Sciences, University of Bristol, Bristol, BS8 1TS, UK,}
\affiliation{School of Chemistry, University of Bristol, Bristol BS8 1TS, UK.}

\author{Jens Eggers}
\affiliation{Department of Mathematics, University Walk, Bristol, BS8 1TW, UK.}

\author{C. Patrick Royall}
\email{paddy.royall@bristol.ac.uk}
\affiliation{H.H. Wills Physics Laboratory, Tyndall Avenue, Bristol, BS8 1TL, UK,}
\affiliation{School of Chemistry, University of Bristol, BS8 1TS, UK,}
\affiliation{Centre for Nanoscience and Quantum Information, Tyndall Avenue, Bristol BS8 1FD, UK.}

\author{Stephen R. Williams}
\affiliation{The Australian National University, Canberra, ACT 0200, Australia}

\author{Hajime Tanaka}
\email{tanaka@iis.u-tokyo.ac.jp}
\affiliation{Institute of Industrial Science, University of Tokyo, 4-6-1 Komaba, Meguro-ku, Tokyo 153-8505, Japan}

\date{\today}

\begin{abstract}
We study the relationship between local structural ordering and dynamical heterogeneities in a model glass-forming liquid, the Wahnstr\"{o}m mixture. A novel cluster-based approach is used to detect local energy minimum polyhedral clusters and local crystalline environments. A structure-specific time correlation function is then devised to determine their temporal stability. For our system, the lifetime correlation function for icosahedral clusters decays far slower than for those of similarly sized but topologically distinct clusters. Upon cooling, the icosahedra form domains of increasing size and their lifetime increases with the size of the domains. Furthermore, these long-lived domains lower the mobility of neighboring particles. These structured domains show correlations with the slow regions of the dynamical heterogeneities that form on cooling towards the glass transition. Although icosahedral clusters with a particular composition and arrangement of large and small particles are structural elements of the crystal, we find that most icosahedral clusters lack such order in composition and arrangement and thus local crystalline ordering makes only a limited contribution to this process. Finally, we characterize the spatial correlation of the domains of icosahedra by two structural correlation lengths and compare them with the four-point dynamic correlation length. All the length scales increase upon cooling, but in different ways. 
\end{abstract}

\maketitle

\section{Introduction}

The nature of the rapid increase in viscosity as liquids are cooled toward the glass transition is the subject of many theoretical approaches, however, there is no consensus on its fundamental mechanism \cite{berthier2011}. One plausible scenario is that dense packing leads to self-induced memory effects, which causes slow dynamics \cite{Gotze2008}. However, the recent discovery of dynamic heterogeneities, i.e., spatial heterogeneities in the relaxation dynamics that emerge on supercooling \cite{hurley1995, ediger2000, berthier}, is suggestive of the importance of a growing dynamic length scale in the slowing down approaching the glass transition. In addition to this dynamical phenomenon, the idea of a structural change leading to vitrification has a long history \cite{frank1952} and recently it has become clear that a range of glass formers show a change in structure upon the emergence of slow dynamics. 

Two generic types of structure have been identified: spatially extendable crystal-like ordering \cite{shintani2006,tanaka2010,sausset2010,leocmach2012}, and non-extendable polyhedral ordering \cite{jonsson1988,kondo1991,tomida1995,julien1996,dzugutov2002,lerner2009,pedersen2010,coslovich2011a}. The latter concerns particles organized into polyhedra \cite{frank1952,tarjus2005} which cannot
 tile Euclidean space due to geometrical frustration \cite{frank1952,tarjus2005}. Thus they form
ramified structures whose fractal dimension is less that the dimensionality of the system, i.e., non-extendable ordering. Metallic glasses can exhibit this second kind of ordering \cite{schenk2002,miracle2004}. Other ``order-agnostic'' schemes have been devised where structural correlations not related to a specific motif have also been identified \cite{biroli2008,mosayebi2010,sausset2011,kob2011non,cammarota2012,reichman2012,dunleavy2012}.

To strengthen the link between structure and dynamics, two approaches have been employed: firstly, dynamically slow regions have been correlated with local ordering. Secondly, 
 emergence of structural and dynamic length scales, which grow similarly, has been sought. The former links structure to dynamical heterogeneity and has been identified in glass formers with both crystal-like ordering \cite{shintani2006,tanaka2010,sausset2010,leocmach2012} and non-extendable polyhedral ordering \cite{dzugutov2002,pedersen2010,coslovich2011a}. 
The latter case of growing structural and dynamic length scales is more controversial.
One of us has identified a direct correspondence between the growing dynamical length scale and a structural length scale in polydisperse hard spheres with crystal-like ordering \cite{tanaka2010,leocmach2012}, while others have not
\cite{charbonneau2012}. In the case of binary Lennard-Jones (Kob-Andersen) mixtures, only weakly growing structural length scales have been identified by a ``point-to-set'' analysis \cite{reichman2012}, while other approaches using static perturbation of inherent structures \cite{mosayebi2010} and finite size scaling \cite{karmakar2012} find a much stronger increase of static length scales. Similarly, for binary hard and soft sphere mixtures, no local ordering, nor one-to-one correspondence between a growing structural length scale and the growing dynamical length scale has been found \cite{widmercooper2006,kob2011non,charbonneau2012,dunleavy2012}, but again static perturbation did suggest a growing structural length scale \cite{mosayebi2012}. 

In some glass formers, local polyhedral ordering in the form of icosahedra has been related to an underlying (quasi-) crystalline structure \cite{pedersen2010,dzugutov1993}. The Wahnstr\"{o}m mixture \cite{wahnstrom1991} used here is a commonly studied binary Lennard-Jones model glass-former of large ($A$) and small ($B$) particles. This model can crystallize upon formation of a structure that contains icosahedra composed of six $A$ and six $B$ species around a central $B$-type particle, which can tile space with the inclusion of Frank-Kasper bonds \cite{pedersen2010}. For this model it was shown that icosahedral order develops upon cooling and a link to slow dynamics has been suggested \cite{dzugutov2002,pedersen2010,coslovich2011a}. 
Thus the Wahnstr\"{o}m mixture provides an intriguing case where icosahedra are found both the in the spatially extendable crystal and in non-extendable networks. In this work, we find the latter.
%It is thus of fundamental importance to reveal how non-extendable icosahedral order is related to dynamical hetereogenities.  

The purpose of this work is threefold: (i) To identify any structural origin for dynamic heterogeneities directly by measuring a structure-specific time correlation function of clusters detected by a topological algorithm.  
(ii) To elucidate the microscopic mechanism by which structured icosahedral domains determine the dynamics, and identify the role that local crystalline ordering plays in this process.
(iii) To understand the behavior of  correlation lengths for the sizes of the structured domains and those for the dynamical heterogeneities.
To address these points, we use the Wahnstr\"{o}m mixture \cite{wahnstrom1991}.

This paper is organized as follows. In section \ref{sectionMethodology}, we provide details of our simulations and analysis, outlining the dynamic topological cluster classification used to elucidate long-lived structural motifs. In our results section \ref{sectionResults}, we show that, of the structures we consider, icosahedral clusters of 13 particles are exceptionally long-lived. These long-lived icosahedra form a network which is correlated with dynamically slow regions, and which acts to slow down neighboring particles. We further show that, although the crystal structure of the Wahnstr\"{o}m mixture is also formed of icosahedra of a specific composition of the A and B particles, these form only a small subset of icosahedra in the supercooled liquid. Finally we show that, of two structural correlation lengths based on these icosahedra, neither scales with the dynamic correlation length $\xi_4(T)$. In Section IV we summarize our work.

\section{Methodology}
\label{sectionMethodology}

\subsection{Model and Simulation Details}

In the Wahnstr\"{o}m mixture \cite{wahnstrom1991}, the two species of Lennard-Jones particles interact with a pair-wise potential,  
$U(r) = 4 \epsilon_{\alpha \beta}[(\frac{\sigma_{\alpha\beta}}{r_{ij}})^{12}-(\frac{\sigma_{\alpha\beta}}{r_{ij}})^6]$, where $\alpha$ 
and $\beta$ denote the atom types $A$ and $B$, and $r_{ij}$ is the separation. The energy, length and mass values are $\epsilon_{AA}=\epsilon_{AB}=\epsilon_{BB}$, $\sigma_{BB}/\sigma_{AA}=5/6$, $\sigma_{AB}/\sigma_{AA}=11/12$ and $m_A=2m_B$ respectively. 
We employ $NVE$ molecular dynamics simulations in 3D with $N=10976$ particles, $N_A=N_B$, at constant density $\rho=1.296$. 
Lengths, temperatures and times are quoted in units of $\sigma_{AA}$, $\epsilon_{AA}/k_B$ and $(m_A \sigma_{AA}^2 / \epsilon_{AA})^{1/2}$ respectively. The molecular dynamics equations of motion are integrated using the Velocity Verlet algorithm with time step $\Delta t=0.001$. The potentials are truncated at $r_c^{AA}=r_c^{AB}=r_c^{BB}=2.5$ and smoothed using the Stoddard-Ford method to ensure continuous forces \cite{stoddard1973}.

\begin{figure}
\centering
\includegraphics[scale=0.65]{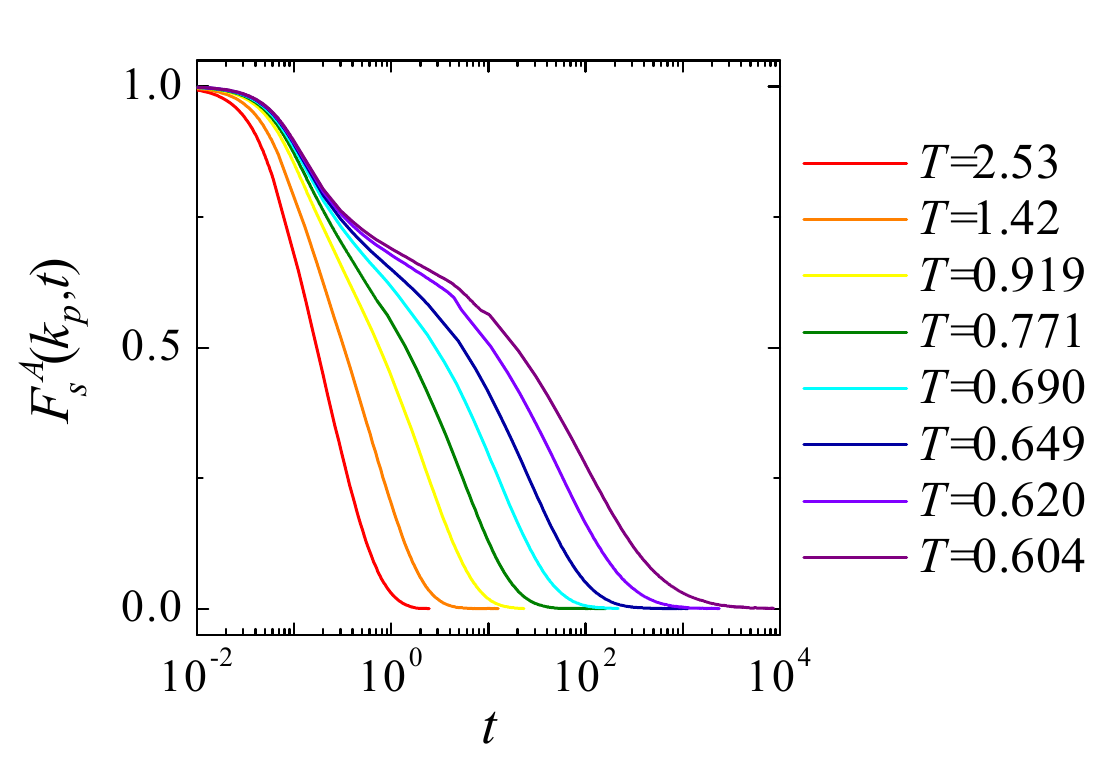}
\caption{The self-intermediate scattering function $F_s^A(k_p,t)$.}
\label{fig:ISF}
\end{figure}

The $\alpha$-relaxation time, $\tau_{\alpha}^{A}$, of each state point is defined using the self-intermediate scattering function (ISF) of the $A$-type particles, $F_{s}^{A}(\textbf{k},t)=\left\langle \sum_{j=1}^{N_{A}}\exp[i\textbf{k}\cdot(\textbf{r}_{j}(t)-\textbf{r}_{j}(0))]\right\rangle$, where $j$ indexes the $A$-species. The angularly averaged $F_{s}^{A}(k_p,t)$, where wavenumber $k_p$ corresponds to the first peak in the $AA$-partial structure factor $S_{AA}(k)$, was fitted with the Kohlrausch-Williams-Watts (KWW) stretched exponential, $F_{s}^{A}(k_p,t)\approx C\exp[-(t/\tau_{\alpha}^{A})^\beta]$. The stretching exponent $\beta$ and Debye-Waller factor $C$ are fitting parameters. Example ISFs are shown in Fig. \ref{fig:ISF}.

The highest temperature state point ($T=5.0$) was initialized from a random configuration. Equilibration took place in the $NVT$-ensemble for $\approx 100\tau_{\alpha}^{A}$ using the Nos\'{e}-Poincar\'{e} thermostat with coupling constant $Q=1.0$ \cite{Nose2001}, before further equilibration in the $NVE$-ensemble for $1000\tau_{\alpha}^{A}$. On completion of the equilibration process, trajectories of length $300\tau_{\alpha}^{A}$ were generated for analysis. 
The state points for $T<5.0$ were obtained via a step-wise cooling process. Each new liquid state point was obtained by quenching instantaneously to a lower temperature from an equilibrated configuration of the previous higher temperature state point. An equilibration process identical to that for the $T=5.0$ liquid was performed following the quench. 
%Trajectories of length $300\tau_{\alpha}^{A}$ were sampled for analysis. 

The stability of the system in the long simulation runs was checked by examining the time evolution of the ISF and the partial radial distribution functions $g_{AA}(r)$, $g_{AB}(r)$ and $g_{BB}(r)$. These quantities do not change throughout the simulation runs. There was also no drift in the numbers of the clusters detected by the TCC algorithm through time. Crystallization was not seen in any of the supercooled samples.

\subsection{The Topological Cluster Classification Method}

\begin{figure}
\centering
\includegraphics[scale=0.48]{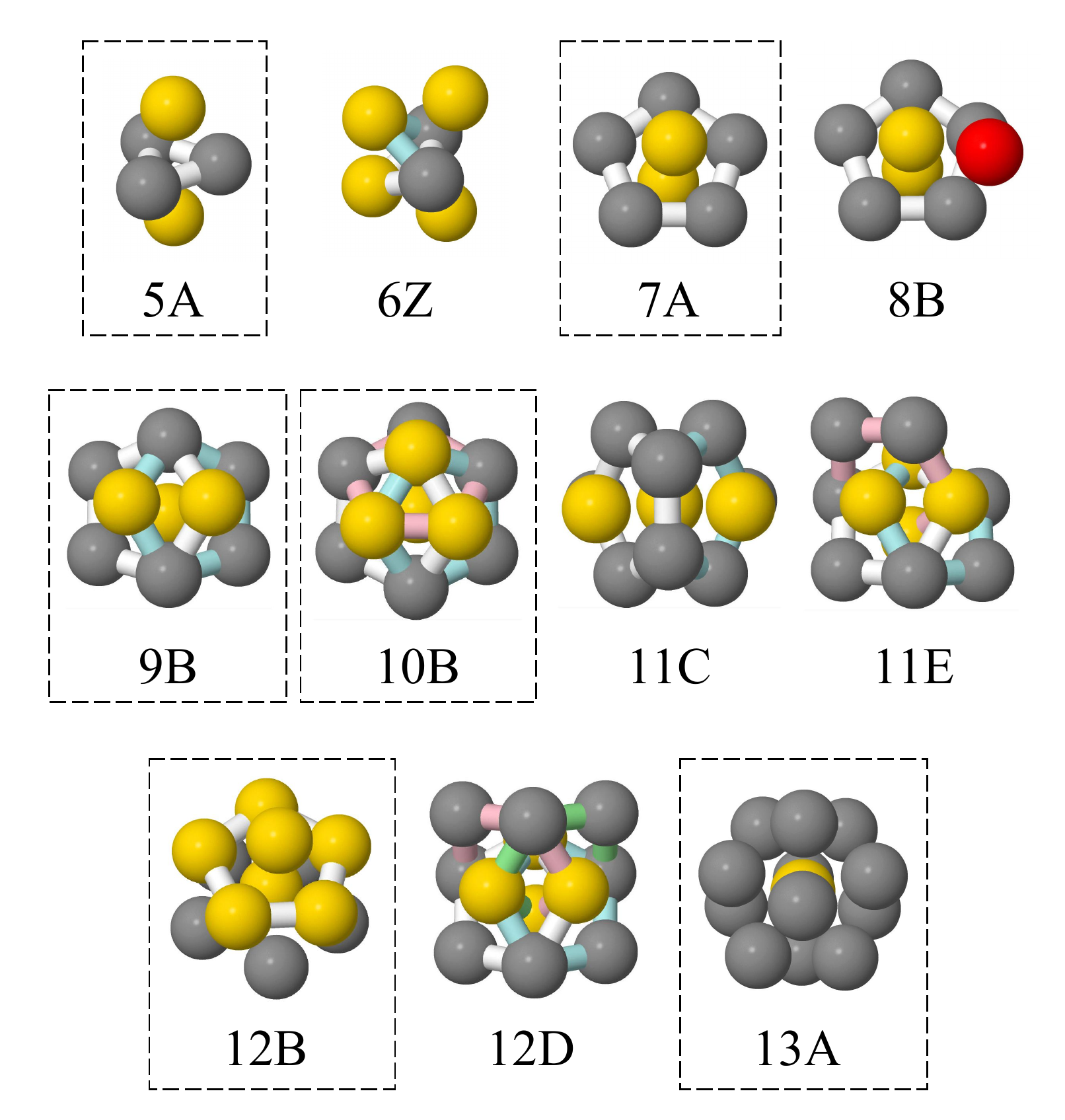}
\caption{The 13A icosahedral cluster and clusters that have icosahedral-like structure. Clusters 11E and 12D occur at $\alpha$-interlocking icosahedral sites \cite{tomida1995}, the rest are contained within a single icosahedron. Clusters in dotted rectangles are ground state clusters of $n$ particles for the Wahnstr\"{o}m mixture \cite{doye2005}. The colors of the particles and the bonds highlight the detection methods in the TCC algorithm. Grey particles are part of 3, 4 or 5 membered shortest path rings. These rings are shown by the white, blue, pink and green colored bonds. Yellow particles are spindle particles for the shortest path rings which together form the basic clusters. Red particles are additional particles bonded to a smaller clusters forming a new cluster. The detection routines for clusters are described in \cite{williams2007}.}
\label{fig:icos_clusts}
\end{figure}

\begin{figure}
\centering
\includegraphics[scale=0.6]{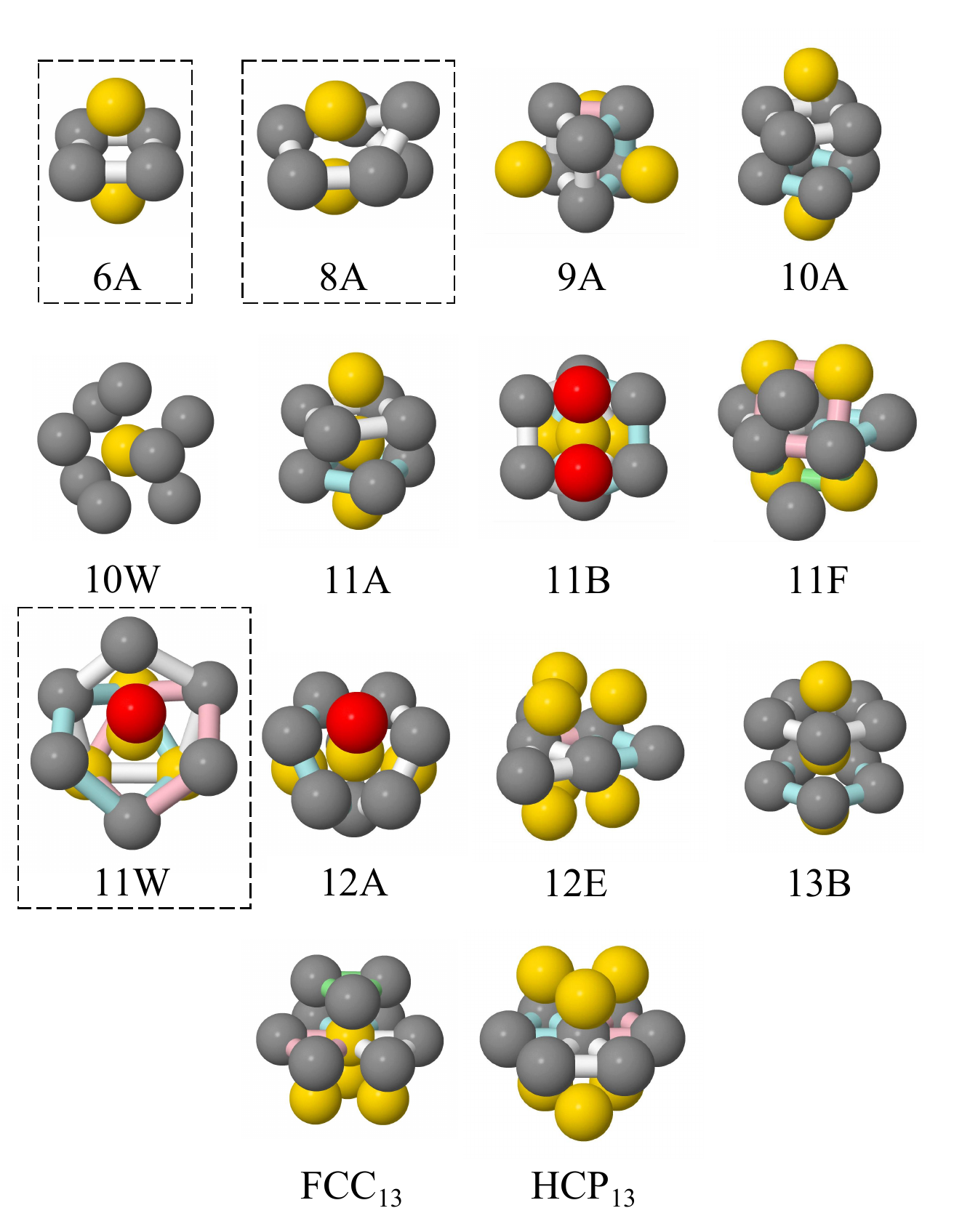}
\caption{Clusters with topology distinct from icosahedral order. Clusters highlighted with dotted rectangles are ground state clusters of $n$ particles for the Wahnstr\"{o}m mixture \cite{doye2005}. Colors are the same as in Fig. \ref{fig:icos_clusts}. Detection routines for 10W and 11W clusters are described below.}
\label{fig:non_icos_clusts}
\end{figure}

Groups of particles whose bond network is topologically identical to isolated clusters which minimize the potential energy \cite{doye2005} are detected using the Topological Cluster Classification algorithm (TCC) \cite{williams2007,royall2008}.  While isolated clusters naturally experience a different environment to that of a bulk (supercooled) liquid, we note that in the case of 13 particles, in a mean-field liquid, the icosahedron has a lower energy than local crystalline ordering \cite{mossa2003}, as is the case for isolated clusters.
The bonds are found using the modified Voronoi construction with four-membered ring parameter $f_c=0.82$. The TCC detects all 3, 4 and 5 membered shortest-path rings within the network of bonded particles, and from these the ground-state clusters are identified. The structure classification is independent of the composition of the cluster in terms of $A$- and $B$- particles. 

We define the lifetime $\tau_\ell$ of a $n$-particle cluster, determined individually by the particle indices, as the time difference between the first and last instances of its detection by the TCC. We ensure that a cluster is detected by the TCC in each $\tau_\alpha^A$ time window during its lifetime $\tau_\ell$ and no subset of the particles become unbonded from the others in the cluster during $\tau_\ell$. The lifetime correlation function $\mathrm{P}(\tau_\ell>t)$ of each type of cluster is the fraction of instances of the cluster that have lifetime $\tau_\ell$ greater than $t$.

The clusters that are identified are all those that are the ground state clusters for the Lennard-Jones potential, the Wahnstr\"{o}m mixture for $n=5$ to $13$ particles, and a range of others that are either ground states of Morse potentials with various interaction lengths, or are clusters found in face-centered cubic (FCC), and hexagonal-close packed (HCP)
crystal structures \cite{williams2007,alexThesis}. 
We follow the nomenclature of Doye \emph{et al.} \cite{doye1995} and divide the clusters into two categories: (i) those that are icosahedral-like in structure, i.e. clusters that are subsets of an icosahedral cluster or occur between sites of $\alpha$- and $\beta$-percolating icosahedral clusters \cite{tomida1995}, and (ii) others with structurally distinct order. The clusters divided between these two categories are depicted in Fig. \ref{fig:icos_clusts} and Fig. \ref{fig:non_icos_clusts} respectively.

For each cluster size $m$, there are $m+1$ different compositions for a cluster in terms of the number of $A$ and $B$-species. The ground-state configurations for each composition may be identified using the publicly available minimization package \texttt{GMIN} \cite{wales1997,GMIN}. While the clusters with the minimum binding energy $E$ for each size $n$ were identified in \cite{doye2005,wales2012}, we list ground state clusters for the other combinations of $n$, $n_A$ in table \ref{tab:ground_states}. 

\begin{table}
\begin{minipage}{1.5in}
  \begin{tabular}{ c | c | c | c }
  	$n$ & $n_A$ & $E$ & TCC \\ \hline
  	5 & 1 & -9.120 & 5A \\
\textbf{5} & \textbf{2} & \textbf{-9.135} & \textbf{5A} \\
5 & 3 & -9.125 & 5A \\
5 & 4 & -9.115 & 5A \\
5 & 5 & -9.104 & 5A \\ \hline
6 & 1 & -12.717 & 6A \\ 
6 & 2 & -12.722 & 6A \\ 
6 & 3 & -12.724 & 6A \\ 
\textbf{6} & \textbf{4} & \textbf{-12.728} & \textbf{6A} \\ 
6 & 5 & -12.719 & 6A \\ 
6 & 6 & -12.712 & 6A \\ \hline
7 & 1 & -16.544 & 7A \\
\textbf{7} & \textbf{2} & \textbf{-16.565} & \textbf{7A} \\
7 & 3 & -16.562 & 7A \\
7 & 4 & -16.539 & 7A \\
7 & 5 & -16.559 & 7A \\
7 & 6 & -16.552 & 7A \\
7 & 7 & -16.505 & 7A \\ \hline
8 & 1 & -19.893 & 8B \\
8 & 2 & -19.964 & 8B \\
8 & 3 & -20.048 & 8A \\
	\end{tabular}
	  \end{minipage}
	%\quad
	\begin{minipage}{1.5in}
	\begin{tabular}{ c | c | c | c }
  	$n$ & $n_A$ & $E$ & TCC \\ \hline
8 & 4 & -20.168 & 8A \\
8 & 5 & -20.206 & 8A \\
\textbf{8} & \textbf{6} & \textbf{-20.233} & \textbf{8A} \\
8 & 7 & -19.995 & 8B \\
8 & 8 & -19.821 & 8B \\ \hline
9 & 1 & -24.206 & 9B \\
9 & 2 & -24.295 & 9B \\
9 & 3 & -24.363 & 9B \\
9 & 4 & -24.429 & 9B \\
9 & 5 & -24.392 & 9B \\
9 & 6 & -24.416 & 9B \\
\textbf{9} & \textbf{7} & \textbf{-24.442} & \textbf{9B} \\
9 & 8 & -24.388 & 9B \\
9 & 9 & -24.113 & 9B \\ \hline
10 & 1 & -28.574 & 10B \\
10 & 2 & -28.713 & 10B \\
10 & 3 & -28.847 & 10B \\
10 & 4 & -28.891 & 10B \\
10 & 5 & -28.871 & 10B \\
10 & 6 & -28.911 & 10B \\
	\end{tabular}
	  \end{minipage}
	%\quad
	\begin{minipage}{2in}
	\begin{tabular}{ c | c | c | c }
  	$n$ & $n_A$ & $E$ & TCC \\ \hline
  	\textbf{10} & \textbf{7} & \textbf{-28.964} & \textbf{10B} \\
10 & 8 & -28.961 & 10B \\
10 & 9 & -28.924 & 10W \\
10 & 10 & -28.423 & 10B \\ \hline
11 & 1 & -32.990 & 11C / 11W \\
11 & 2 & -33.168 & 11C / 11W \\
11 & 3 & -33.310 & 11C / 11W \\
11 & 4 & -33.426 & 11C / 11W \\
11 & 5 & -33.477 & 11C / 11W \\
11 & 6 & -33.604 & 11W \\
11 & 7 & -33.770 & 11W \\
11 & 8 & -33.917 & 11W \\
\textbf{11} & \textbf{9} & \textbf{-33.967} & \textbf{11W} \\
11 & 10 & -33.937 & 11W \\
11 & 11 & -32.766 & 11C / 11W \\ \hline
12 & 1 & -38.279 & 12B \\
12 & 2 & -38.568 & 12B \\
12 & 3 & -38.772 & 12B \\
12 & 4 & -38.893 & 12B \\
12 & 5 & -38.988 & 12B \\
	\end{tabular}
	  \end{minipage}
	%\quad
	\begin{minipage}{1.5in}
		\begin{tabular}{ c | c | c | c }
  	$n$ & $n_A$ & $E$ & TCC \\ \hline
12 & 6 & -38.991 & 12B \\
12 & 7 & -39.040 & 12B \\
\textbf{12} & \textbf{8} & \textbf{-39.056} & \textbf{12B} \\
12 & 9 & -39.039 & 12B \\
12 & 10 & -38.944 & 12B \\
12 & 11 & -38.841 & 12B \\
12 & 12 & -37.968 & 12B \\ \hline
13 & 1 & -44.583 & 13A \\
13 & 2 & -44.826 & 13A \\
13 & 3 & -45.076 & 13A \\
13 & 4 & -45.261 & 13A \\
13 & 5 & -45.391 & 13A \\
13 & 6 & -45.528 & 13A \\
13 & 7 & -45.537 & 13A \\
\textbf{13} & \textbf{8} & \textbf{-45.554} & \textbf{13A} \\
13 & 9 & -45.531 & 13A \\
13 & 10 & -45.475 & 13A \\
13 & 11 & -45.411 & 13A \\
13 & 12 & -45.374 & 13A \\
13 & 13 & -44.327 & 13A \\
\end{tabular}
  \end{minipage}
  \caption{The ground state clusters of the Wahnstr\"{o}m mixture in TCC notation for $n=5$ to $13$ particles containing $n_A$ $A$-specie particles. The energy $E$ is the binding energy of the cluster. Clusters highlighted in \textbf{bold face} are the lowest energy states for a given value of $n$. The final column is the cluster detected by the Topological Cluster Classification algorithm from the ground state configuration. Note that for the Wahnstr\"{o}m mixture the ground state clusters with $n_A=0$ and $n_A=n$ are identical for each $n$ as the $AA$ and $BB$ Lennard-Jones interactions are the same. The $n_A=0$ clusters have therefore been omitted from the table for brevity. }
  \label{tab:ground_states}
\end{table}

The detection of structures by TCC is independent of the type of species present in the cluster, except for when icosahedral clusters compatible with the crystal phase are considered below. Table \ref{tab:ground_states} demonstrates that the structure of the ground state clusters of the Wahnstr\"{o}m mixture for each size $n$ is only weakly dependent on the composition of the particles. The TCC algorithm presented in \cite{williams2007,royall2008} is able to identify the structure of all but two clusters in terms of the Morse ground state clusters. The additional two clusters 10W and 11W are specific to the Wahnstr\"{o}m mixture.

The cluster 10W is composed of six five-membered rings of particles forming a shell around a central particle. All particles in the shell are bonded to the central particle, and the central particle is the only nearest neighbor shared by all particles in each of the five-membered rings. This is the ground state for the scenario $n=10$ and $n_A=9$.
For $n=11$ there are multiple compositions where the 11W cluster is the ground state. This is identified as a 10B cluster with a single additional particle bonded to the central 10B particle. The coordination number is 10.
The crystal structure of the Wahnstr\"{o}m mixture contains icosahedral clusters and Frank-Kasper clusters of a specific composition. Their relationship to the icosahedral domains is discussed in the next section.

\section{Results and discussion}

\label{sectionResults}

\subsection{Overall dynamics}

\begin{figure}
\includegraphics[scale=0.85]{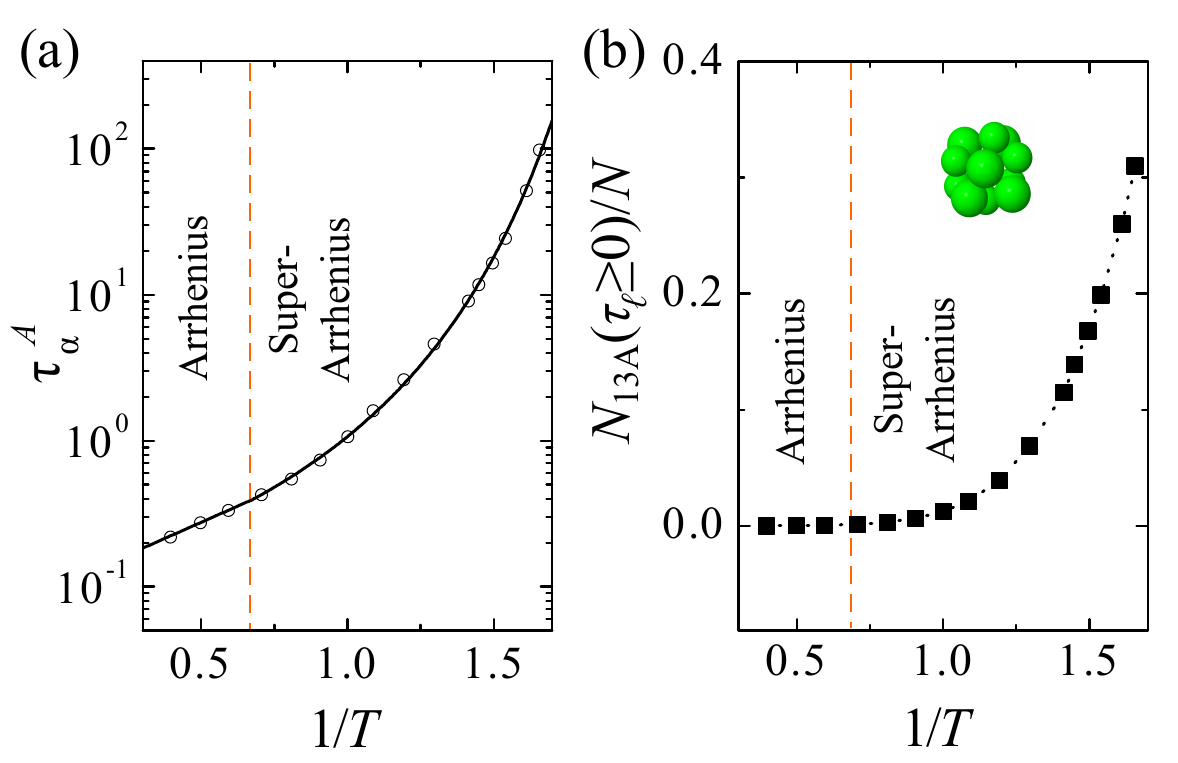}
\caption{(a) Non-Arrhenius increase in structural relaxation time $\tau_\alpha^A$ versus the inverse of temperature. The data are fitted with an Arrhenius exponential at high temperatures and VFT at low temperatures. The onset temperature of the super-Arrhenius behavior is $T^\ast=1.48$ (vertical dashed line in both (a) and (b)). (b) The fraction of particles detected within icosahedral 13A clusters of all lifetimes $N_{13\mathrm{A}}(\tau_\ell\ge0)/N$. Note the fast increase in $N_{13\mathrm{A}}(\tau_\ell\ge0)/N$ starts around $T^\ast$.}
\label{fig:nonArrhen}
\end{figure}

First we show the temperature dependence of the structural relaxation time $\tau_\alpha$ in Fig. \ref{fig:nonArrhen}(a).
As is usual for fragile glass-forming systems, we see two regimes for the behavior of the structural relaxation time delimited by an onset temperature for slow dynamics $T^\ast=1.48$. 
For $T>T^\ast$ an Arrhenius form, $\tau_{\alpha}^{A} = \tau_{\infty} {\exp} [ E_{\infty} / k_B T]$, where $\tau_{\infty}$ and $E_{\infty}$ are fitting parameters, describes $\tau_{\alpha}^{A}(T)$ well. For $T\le T^\ast$ the Vogel-Fulcher-Tammann (VFT) form is used, $\tau_{\alpha}^{A} = \tau_{\infty}^{'} \exp [ DT_0/ (T-T_{0})]$, where $D$ is the fragility index and $T_0=0.40$, for the non-Arrhenius regime of increasing relaxation times. 
Hereafter we seek structural features responsible for the super-Arrhenius behavior. 

\subsection{Topological Cluster Classification analysis}

The temperature dependence of the number density of icosahedral structures is shown in Fig. \ref{fig:nonArrhen}(b). The fraction of particles detected as members of icosahedral clusters is $N_{13\mathrm{A}}(\tau_\ell\ge0)/N$. Here $\tau_\ell\ge0$ emphasizes that here we consider all icosahedra rather than the time-correlated structure considered below.
At high temperatures there are few icosahedral clusters. On cooling there is a rapid increase in $N_{13\mathrm{A}}(\tau_\ell\ge0)/N$. This coincides with the onset of the super-Arrhenius increase in $\tau_\alpha^A$ [see Fig. \ref{fig:nonArrhen}(a)], suggesting a link between structure and dynamics in this fragile system. For the lowest temperature sample, $T=0.604$, approximately 30\% of particles are found to be within icosahedral clusters.

\begin{figure}
\centering
\includegraphics[scale=0.75]{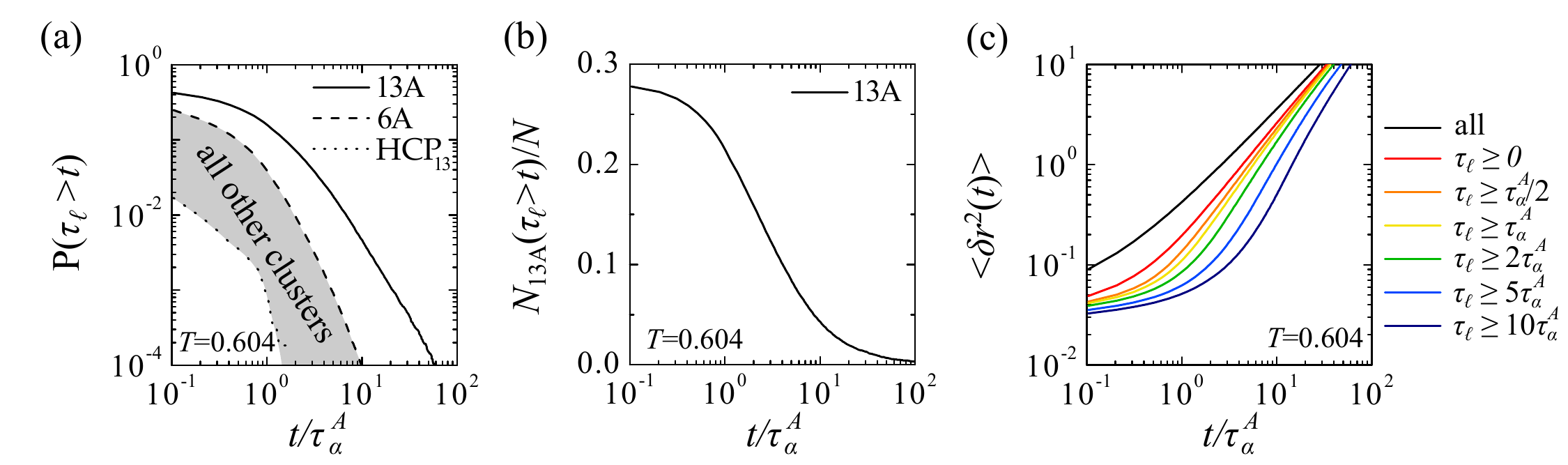}
\caption{(a) Cluster lifetime correlation functions $\mathrm{P}(\tau_\ell>t)$ for icosahedral (denoted 13A - solid line), octahedral (6A - dashed line), and HCP$_{13}$ (dotted line) at $T=0.604$. All clusters with structure distinct from icosahedral order fall in the grey shaded region between $\mathrm{P}(\tau_\ell>t)$ for 6A and HCP$_{13}$ . (b) The fraction of particles detected within icosahedral clusters with lifetime $\tau_\ell>t$ ($T=0.604$). (c) The mean-square displacement $\langle \delta r^2(t) \rangle$ of all particles (black line) and particles starting within icosahedral clusters with lifetime $\tau_\ell>t$ as specified in the legend (colored lines - $T=0.604$).}
\label{fig:cluster_lifetime}
\end{figure}

In Fig. \ref{fig:cluster_lifetime}(a) the cluster lifetime correlation functions $\mathrm{P}(\tau_\ell>t)$ are shown for a range of clusters detected by the TCC in the lowest temperature sample ($T=0.604$). The lifetime correlation function for the $n=13$ icosahedral clusters (black line) decays much more slowly than $\mathrm{P}(\tau_\ell>t)$ for all the structurally distinct clusters detected by the TCC (grey shaded region).

The long-time tail of the icosahedral correlation function indicates that some of these clusters preserve their local structure on timescales far longer than $\tau_{\alpha}$. 
This effect may be related to previous observations that some icosahedra (those in which the central particle was previously in an icosahedron) are more stable than average \cite{kondo1991}.
As we shall see below, this effect is enhanced when the icosahedra group into domains.
The lifetimes of all cluster types hold no simple monotonic relationship to their size and frequency of occurrence. For example the four-fold symmetric $n=6$ particle octahedral cluster is smaller and occurs 8 times more frequently than icosahedral clusters, yet displays faster decay of $\mathrm{P}(\tau_\ell>t)$. Moreover, there is no trend in the rate of decay of the lifetime correlation function $\mathrm{P}(\tau_\ell>t)$ for the ground-state clusters, even though it was once conjectured that the ground-state clusters would be most stable \cite{frank1952}. For example, the ground state clusters 6A, 8A and 11W have correlation functions $\mathrm{P}(\tau_\ell>t)$ that decay more quickly than the ground-state clusters that are subsets of the icosahedron (5A, 7A, 8B, 9B, 10B, 12B and 13A). The correlation functions $\mathrm{P}(\tau_\ell>t)$ for the 6A, 8A and 11W clusters fall within the shaded region in Fig. \ref{fig:cluster_lifetime}(a).

The fast initial drop in $\mathrm{P}(\tau_\ell>t)$ reflects the existence of large numbers of icosahedra with lifetime $\tau_\ell \ll \tau_\alpha^A$. The lifetime of these clusters is comparable to the timescale for the beta-relaxation regime where the 
particles fluctuate within their `cage' of neighbors. These clusters are not representative of the underlying viscous structure of the liquid that exists on the timescales of the dynamic heterogeneities $\approx \tau_\alpha^A$. Fig. \ref{fig:cluster_lifetime}(b) demonstrates that particles within these short-lived icosahedral clusters make only a small additional contribution to the total fraction of particles detected within all icosahedral clusters. Icosahedral clusters overlap and particles can be members of multiple icosahedral clusters simultaneously, and the results of Fig. \ref{fig:cluster_lifetime}(b) indicate that short- and long-lived icosahedra mainly lie in the same regions of the liquid. Here we do not distinguish the short- and long-lived icosahedral clusters structurally.

The mean-square displacement (MSD), $\langle \delta r^2(t) \rangle=1/N \sum_{i}|\mathbf{r}_i(t)-\mathbf{r}_i(0)|^2$ is shown in Fig. \ref{fig:cluster_lifetime}(c) for all particles (solid black line) and for particles initially within icosahedral clusters (colored lines). The dynamics of particles within icosahedra of all lifetimes are slower than the system-wide average (c.f. black line). The longer the lifetime of the icosahedra, the slower the particles become. 
Since some icosahedra last for very long terms [Fig. \ref{fig:cluster_lifetime}(b)],
we expect that these particles may exhibit very low mobilities.

\begin{figure}
\includegraphics[scale=0.9]{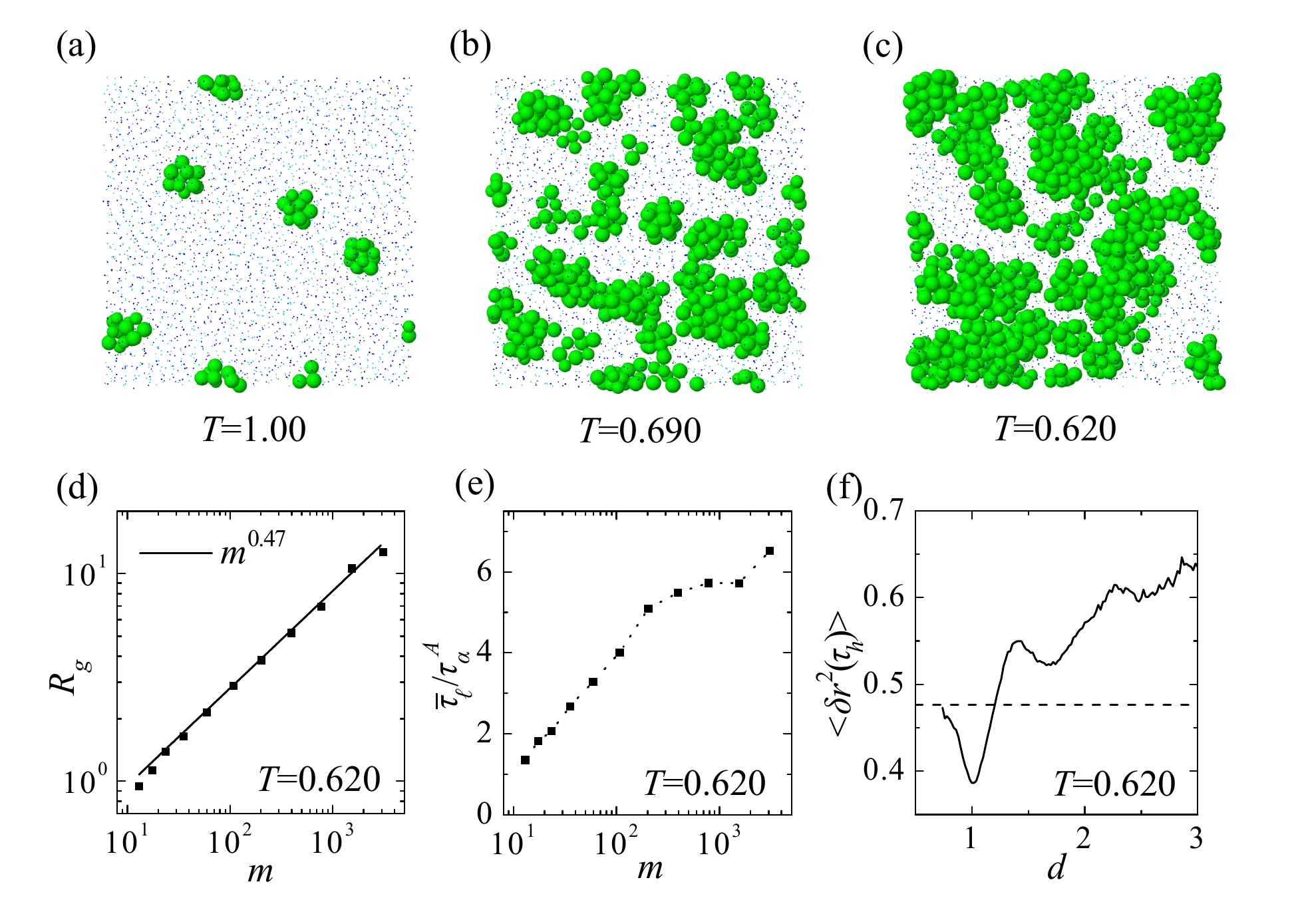}
\caption{(a)-(c) Domains of icosahedral clusters form on cooling from high to low temperature (slices through 3D simulation box). Particles in icosahedral clusters are shown full size in green, other particles are blue dots. (d) The radius of gyration $R_g$ of the domains versus the number of particles in the domain $m$ for $T=0.620$. $R_g$ is well fitted by $m^{0.47}$ indicating the domains have a fractal dimension $d_f \simeq 2$. (e) The mean lifetime of icosahedral clusters $\bar{\tau_\ell}$ versus the domain size $m$. (f) 
Icosahedral domains retard the motion of neighboring particles. The MSD 
$\langle \delta r^2(\tau_h) \rangle$ as a function of distance from icosahedral domains $d$ (solid line).  The dotted line is the MSD over $\tau_h$ of all particles not in icosahedra, independent of $d$. Note that the particles located around $d=1$ amount to 40\% of the system at this temperature. }
\label{fig:frac_dim}
\end{figure}

\subsection{Network of icosahedra}

The domains of icosahedra that form on cooling are shown in Fig. \ref{fig:frac_dim}. At high temperature icosahedra are predominantly isolated [Fig. \ref{fig:frac_dim}(a)]. Upon cooling, the icosahedra overlap and join together [Fig. \ref{fig:frac_dim}(b)] to form (transient) networks at low temperatures [Fig. \ref{fig:frac_dim}(c)]. In order to investigate the structure of the domains, we calculate the radius of gyration $R_g$, $R_g^2=\frac{1}{2m^2} \sum_{i,j}(\mathbf{r}_i-\mathbf{r}_j)^2$, where $m$ is the number of particles in the domain and the sum extends over all pairs of particles in the domain. The radius of gyration shows power-law growth with an exponent of $0.47$ with the size of the domain $m$ [Fig. \ref{fig:frac_dim}(d)]. In other words the domains have a fractal dimension $d_f \simeq 2$, indicating that they are non-space-filling. We find that the individual icosahedra have enhanced stability as the size of the icosahedral domains grow [Fig. \ref{fig:frac_dim}(e)]. The time $\bar{\tau_\ell}$ is the mean lifetime of icosahedra taken from size $m$ domains in each configuration. For $T=0.620$, this grows almost five-fold as the domain size increases from isolated icosahedra ($m=13$) to extended clusters of icosahedra 
($m \approx 200$). 
For higher $m$, there is some suggestion that $\bar{\tau_\ell}$ may saturate.
At present, it is hard to be sure. Our data are limited by system size, so much larger simulations would be required to properly address this issue. If indeed, there is saturation in $\bar{\tau_\ell}$ with $m$, this would suggest a maximal stability of clusters. Thus that upon percolation where $m$ diverges, $\bar{\tau_\ell}$ need no increase further and full dynamical arrest would not be expected, as we indeed find.

\begin{figure}
\includegraphics[scale=0.95]{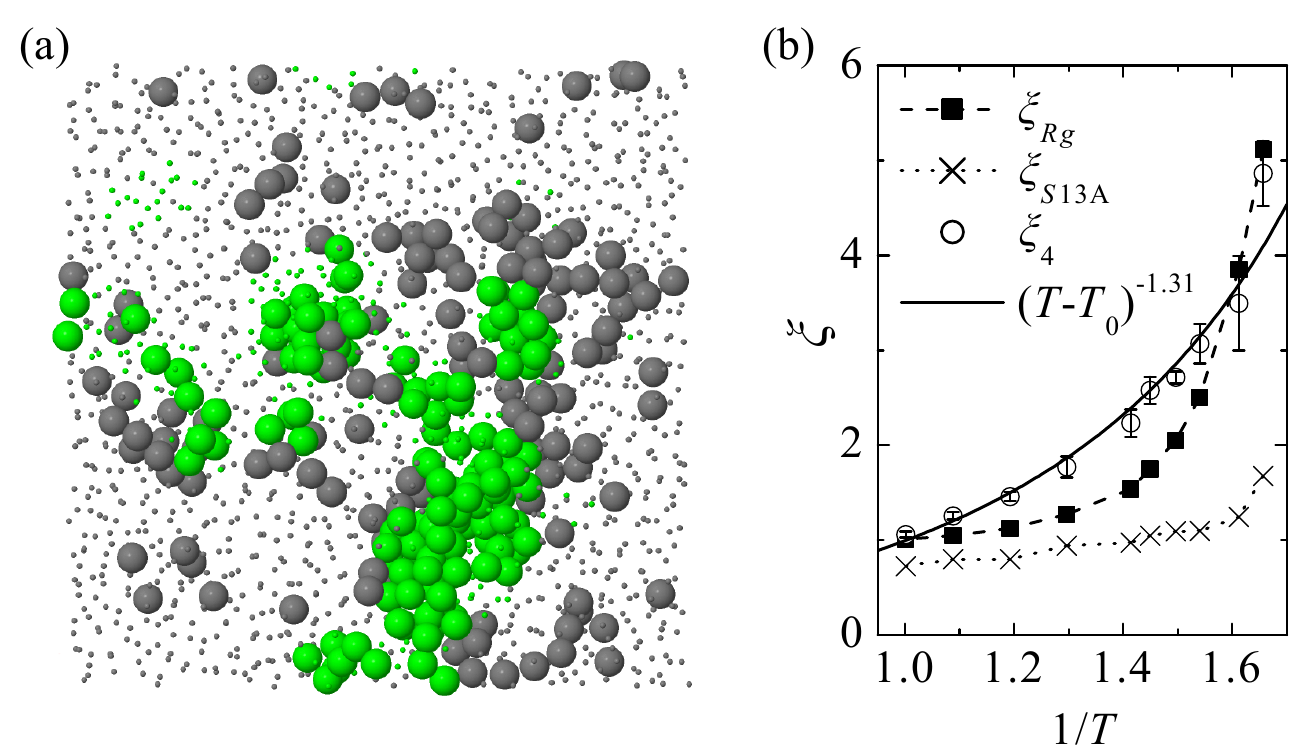}
\caption{(a) The relationship between icosahedra and spatial heterogeneities in the dynamics at $T=0.604$ (2D slice of simulation box). Mobility ($\langle \delta r^2(\tau_h) \rangle$) is depicted by the size of the particle. Large particles have $\langle \delta r^2(\tau_h) \rangle < 0.043$, small particles otherwise. Particles always within icosahedral clusters with lifetime $\tau_\ell \ge \tau_\alpha^A$ are colored green, and others are grey. (b) Growing structural $\xi_{Rg}$ (squares) and $\xi_{S13\mathrm{A}}$ (crosses), and dynamical  $\xi_{4}$ (circles) correlation lengths.}
\label{fig:dyn_het}
\end{figure}

We now consider the effect this network has on the remainder of the system. In  Fig. \ref{fig:frac_dim}(f), we plot the mean-square displacement of the non-icosahedral particles $\langle \delta r^2(\tau_h) \rangle$ against the distance $d$ from a domain of icosahedra at time $t=0$. Here the timescale is taken as the maximum of the four-point dynamic susceptibility $\chi_4(t)$ \cite{berthier2011}, $\tau_h \simeq \tau_{\alpha}^A$. The first nearest neighbors of the domains have suppressed mobility, due to coupling to the stable and slow domains. The second and third shells of neighbors of the domains have higher mobility. This connection between the domains of icosahedra and the spatially heterogeneous dynamics is depicted in Fig. \ref{fig:dyn_het}(a). The transient network of slow icosahedral particles retards the dynamics of neighboring particles, leading to a mechanism for the formation of dynamical heterogeneities. 
This is reminiscent of a point-to-set mechanism \cite{biroli2008,cammarota2011} whereby pinned (immobilized) particles inhibit the motion of their surroundings. Of course here the particles in long-lived icosahedra are slow, rather than immobilized, so the network is transient, but its neighbors are nevertheless retarded.

\subsection{Relationship of icosahedral domains to crystalline order}

\begin{figure}
\centering
\includegraphics[scale=0.85]{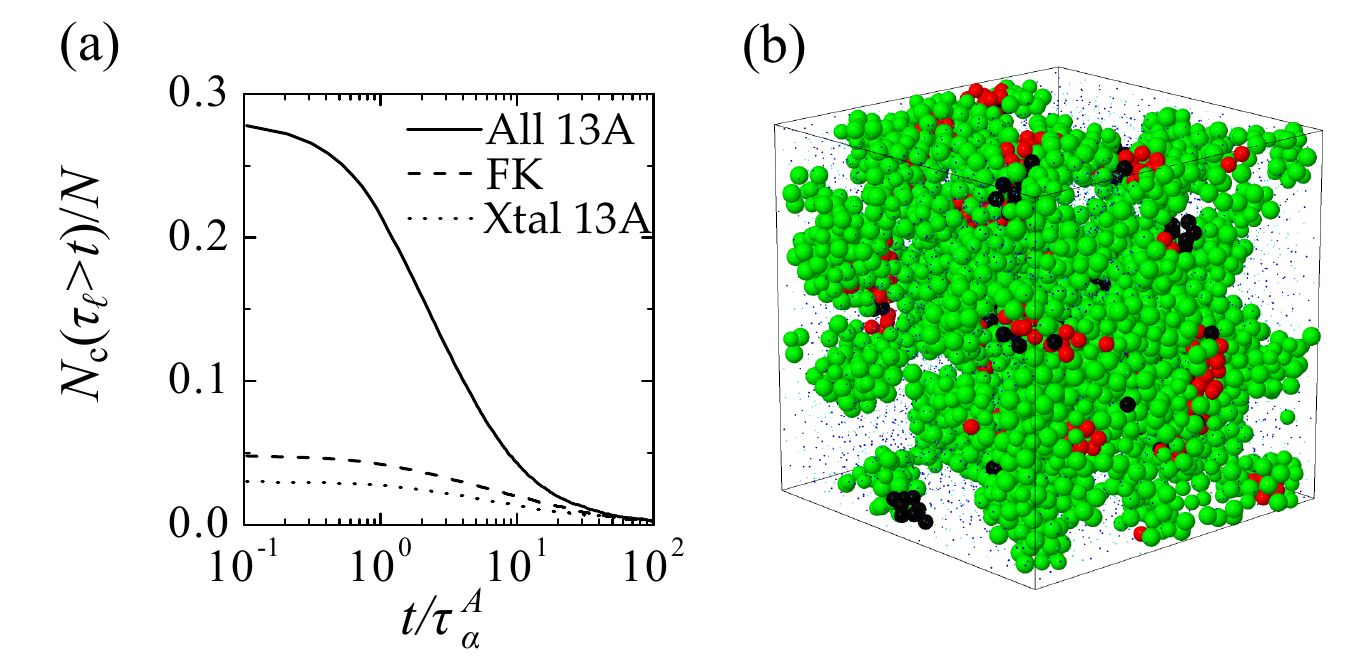}
\caption{(a) Relationship between icosahedral 13A domains and the crystalline `Frank-Kasper' and icosahedral clusters for $T=0.604$. The quantity $N_{\mbox{c}}(\tau_\ell>t)/N$ is the fraction of particles detected within clusters lifetime $\tau_\ell > t$. The dashed line is for Frank-Kasper clusters, and the dotted line is for icosahedral clusters with a $B$-particles in the center and arrangements of $A$- and $B$-species in the shell of the icosahedron compatible with the bulk crystal phase \cite{pedersen2010}. Both of these clusters exist in the MgZn$_2$ structure of the Wahnstr\"{o}m crystal. The solid line is for all icosahedra, irrespective of the arrangement of $A$- and $B$-species. (b) Snapshot at $T=0.604$ showing the overlap between icosahedral domains and Frank-Kasper clusters. Green particles are the icosahedral domains, red particles are those in both Frank-Kasper and icosahedral clusters, and black are Frank-Kasper particles \textit{not} in icosahedral domains. The small blue particles are not members of icosahedral or Frank-Kasper clusters.}
\label{fig:gCryst13AFK}
\end{figure}

Pedersen \textit{et al.} identified that the crystal structure of the Wahnstr\"{o}m mixture has the structure of the MgZn$_2$ Laves phase, which is formed of 13A icosahedra and Frank-Kasper bonds with specific compositions arrangements of the $A$- and $B$-species\cite{pedersen2010}. They found that icosahedra within the crystal structure are composed of six of the larger $A$-particles and seven of the smaller $B$-species. Furthermore of all the possible arrangements of the $A$- and $B$-particles within the icosahedral cluster for this composition, only two are present in the crystal structure. They also suggested that Frank-Kasper clusters, consisting of two $A$-specie particles bonded to six common $B$-specie neighbors, that occur in the crystalline lattice, act to stabilize the supercooled liquid. Here we discuss the relationship between the icosahedral domains we identify and crystalline structure in the supercooled liquid.

In Fig. \ref{fig:gCryst13AFK}(a) we present data for the fraction of particles detected within clusters of lifetime $\tau_\ell > t$ for three types of clusters. First are for the icosahedral clusters (13A) that form extend domains through the system. These clusters can be made of any number $A$- and $B$-species, i.e. icosahedra with any compositions of $A$- and $B$-particles and any arrangement of the particles within the cluster for a given composition. Second are the icosahedral clusters compatible with the crystal structure \cite{pedersen2010}. There are two different arrangements that consist of six $A$-particles and six $B$-particles arranged around a central $B$-particle. These are identified by the number of connected $B$-species in two sets on the shell of the icosahedral cluster which are separated by a ring of six $A$-particles. One arrangement has two sets of three connected $B$-particles, and the other one pair and one quadruplet of $B$-particles. The particles within crystal-compatible icosahedral clusters are necessarily a subset of the particles within all icosahedral The third cluster is the Frank-Kasper cluster, which consists of a bonded pair of $A$-species surrounded by a ring of 6 $B$-particles. 

For all cluster lifetimes there are more particles in icosahedral clusters than in Frank-Kasper or crystal compatible icosahedra. It is strictly necessary that particles in crystal compatible icosahedral clusters are fewer than in icosahedral clusters, as the former are by definition icosahedra themselves. However we emphasize that very few particles ($<5\%$) are identified in the clusters compatible with the crystal for the lowest temperature state point we studied.
We find for $T=0.604$ that 83\% of particles in Frank-Kasper clusters also lie within the icosahedral domains. The overlap of particles in Frank-Kasper clusters and icosahedral domains is shown in Fig. \ref{fig:gCryst13AFK}(b).

We conclude that crystalline structuring in supercooled Wahnstr\"{o}m liquids is a relatively small component of wider icosahedral ordering. The icosahedral ordering is frequently incompatible with the crystal structure due to the composition, i.e. $n_A$ is frequently not equal to 6, or because the arrangements of the $A$- and $B$-species within the cluster are inconsistent with the crystalline icosahedra.

\subsection{Comparison of structural and dynamic length scales}

Finally we investigate how the four-point correlation length for the dynamic heterogeneities $\xi_4$ compares to two static correlation lengths for the sizes of the domains of icosahedra.  The first of these is the radius of gyration of the icosahedral clusters $\xi_{Rg}=  R_g^{13\mathrm{A}} (\langle m \rangle / 13)^{0.47}$, where $\langle m \rangle$ is the ensemble average of the domain size and $R_g^{13\mathrm{A}}=0.95$ is the radius of gyration of a single icosahedral cluster. This demonstrates the growth in the domains on cooling. The second, which we term $\xi_{S13\mathrm{A}}$ which is calculated from fitting the Ornstein-Zernike (OZ) equation [Eq. \ref{eqOZ}] to the low $k$ behavior of the static structure factor restricted to the icosahedral particles $S_{13\mathrm{A}}(k)$.

We follow La\v{c}evi\'{c} \textit{et al}. in our calculation of the dynamic correlation length $\xi_{4}$ \cite{lacevic2003}. We define an overlap function, 
$w(|\textbf{r}_j(0)-\textbf{r}_l(t)|)$, to be unity if $|\textbf{r}_j(0)-\textbf{r}_l(t)|\le a$, 0 otherwise, where $a=0.3$. It defines the particle sites at time $0$ that are occupied by a particles at time $t$. The fraction of overlapping particles between two configurations is 

\begin{equation}
Q(t)=\frac{1}{N} \sum_{j=1}^N \sum_{l=1}^N w(|\textbf{r}_j(0)-\textbf{r}_l(t)|). 
\label{eqQ}
\end{equation}

\begin{figure}
\centering
\includegraphics[scale=0.75]{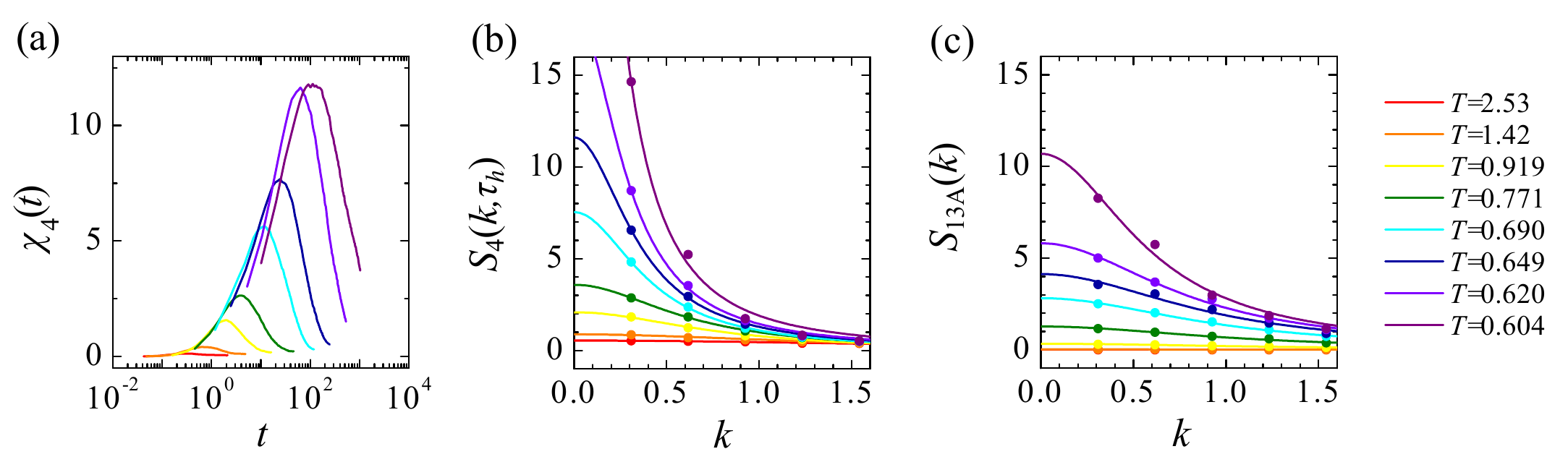}
\caption{(a) The four-point dynamic susceptibility $\chi_4(t)$. The maximum of $\chi_4(t)$ occurs at time $\tau_h$. (b) The four-point dynamic structure factor $S_4(k,\tau_h)$. The solid lines are fits to the data with the  Ornstein-Zernike equation [\ref{eqOZ}]. (c) The structure factor of the icosahedral particles $S_{13A}(k)$. The solid lines are fits to the data with the OZ function.}
\label{fig:S4}
\end{figure}

The four-point dynamic susceptibility $\chi_4(t)$ is then given by 
\begin{equation}
\chi_4(t)=\frac{V}{N^2 k_B T}[\langle Q(t)^2\rangle -\langle Q(t)\rangle^2]. 
\label{eqChi4}
\end{equation}
The time at which $\chi_4(t)$ has a peak is $\tau_h$ [Fig. \ref{fig:S4}(a)].

The four-point dynamic structure factor $S_4(\textbf{k},t)$ is defined as
\begin{equation}
S_4(\textbf{k},t)=\frac{1}{N\rho} \langle \sum_{jl} \exp[i \textbf{k} \cdot \textbf{r}_j(0)]w(|\textbf{r}_j(0)-\textbf{r}_l(t)|) \sum_{mn} \exp[i \textbf{k} \cdot \textbf{r}_m(0)]w(|\textbf{r}_m(0)-\textbf{r}_n(t)|) \rangle,
\label{eqS4}
\end{equation} 
where $j$, $l$, $m$, $n$ are particle indices. For time $\tau_h$, the angularly averaged version is $S_4(k,\tau_h)$. The dynamic correlation length $\xi_4$ is calculated by fitting the OZ equation, 
\begin{equation}
S_4(k,\tau_h)=\frac{S_4(0,\tau_h)}{1+[k\xi_4(\tau_h)]^2}
\label{eqOZ}
\end{equation}
to $S_4(k,\tau_h)$ for $k<2$ \cite{lacevic2003} [Fig. \ref{fig:S4}(b)]. To calculate $\xi_{S13A}$, we follow a similar approach as for $\xi_4$. The structure factor of the icosahedral particles is given by 

\begin{equation}
S_{13A}(\textbf{k})=\frac{1}{N\rho} \langle \sum_{j=1}^{N_{13A}} \sum_{l=1}^{N_{13A}} \exp[-i \textbf{k} \cdot \textbf{r}_j(0)]\exp[i \textbf{k} \cdot \textbf{r}_l(0)] \rangle,
\label{eqS13A}
\end{equation} 
where $j$, $l$ index particles that are detected within an icosahedral cluster and $N_{13A}=N_{13A}(\tau_\ell\ge0)$ is the total number of icosahedral particles in the configuration. The correlation length $\xi_{S13A}$ is found by fitting the Ornstein-Zernike (OZ) equation to the low $k$ values of the radially averaged icosahedral structure factor $S_{13A}(k)$ [Fig. \ref{fig:S4}(c)].

The temperature behavior of these lengths is shown in Fig. \ref{fig:dyn_het}(b).
All three correlation lengths increase on cooling. 
For the lowest temperature studied, the network of icosahedra percolates, thus
 $\xi_{Rg}$ cannot be defined, suggesting a divergence in a structural length scale at a temperature higher than that inferred from the dynamic correlation length $\xi_{4}$. Thus percolating domains of icosahedra do not imply structural arrest (see the discussion in Sec. III.C): since icosahedra have a finite lifetime, the formation of a network does not lead to arrest, unlike the case of colloidal gels \cite{royall2008}. The length $\xi_{S13 \mathrm{A}}$ grows more slowly than $\xi_{4}$ as the domains of icosahedra are smaller than the immobile regions selected by the four-point dynamic susceptibility $\chi_4(t)$ [see Fig. \ref{fig:dyn_het}(a)]. 
This result is expected because domains of icosahedra in this liquid form rarefied networks with $d_f \simeq 2$ which do not fill space.
It is unlikely that this difference would be rectified by including the immediate neighbors of icosahedral domains when calculating the structural correlation lengths, because there are large numbers of neighbors of icosahedral domains that are found at the lowest temperatures. We also find that there are a fraction of the particles within icosahedral clusters that are not selected in the calculation of $\xi_4$ (i.e. their displacement over $\tau_h$ is greater than 0.3, meaning that there is not a perfect correlation between the structured and the slow particles.
Note $\chi_4$ has been shown to exhibit dependence upon system size for $N \lesssim 1000$ \cite{karmakar2009}. We expect that such effects are reasonably small here, and have taken care to only consider temperatures where all our measured lengths are much smaller than the system size. While system size effects cannot be ruled out, we do not believe these make a significant impact on the conclusions we draw.

The result for this Wahnstr\"{o}m mixture contrasts with liquids with crystal-like order. 
In some work, this has been found to scale with $\xi_4$, \cite{shintani2006, tanaka2010, leocmach2012}. Other methods have also shown that structural correlation lengths are rather
 larger in the case of such liquids \cite{dunleavy2012}.
%
%
% whose structural correlation lengths have been found to scale with $\xi_4$ due to the space-
%illing nature of the locally preferred structure. 
We emphasize that the icosahedral clusters which form the 
domains have arrangements of the $A$- and $B$-species that are rarely compatible with the crystal structure. Only 10\% of the particles within icosahedral domains are members of clusters in the crystal structure. Thus, 
although an icosahedral structure itself is a key structural motif of the Frank-Kasper crystal, icosahedral clusters formed in a supercooled liquid have a rather weak link to the crystal structure [Fig. \ref{fig:gCryst13AFK}(b)], presumably due to entropic frustration \cite{pedersen2010} of the chemical arrangement of the species within the cluster.  

The difference in the nature of the structural fluctuations between this model and those with spatially extendable crystal-like order may be responsible for the difference in the $T$-dependence of $\xi_4$: $\xi_4 \propto (T-T_0)^{-1.31}$ is found here whereas 
$\xi_4 \propto (T-T_0)^{-2/3}$ for liquids with extendable crystal-like \cite{shintani2006,tanaka2010,leocmach2012} or amorphous order \cite{mosayebi2010,mosayebi2012}.  
Thus, the question remains as to the most appropriate structural correlation length to explain the viscous slowing down of supercooled liquids, and how order-specific correlation lengths, such as $\xi_{Rg}$ and $\xi_{S13\mathrm{A}}$, might relate to order-agnostic structural correlation lengths \cite{biroli2008,sausset2011,dunleavy2012,cammarota2012}. We also note that Fig. \ref{fig:frac_dim}(f) strongly indicates that the effect of the domains on the surrounding particles extends around $\sigma$ from the domains, showing that the dynamical effect of icosahedral clusters is not just limited to their domains. Thus icosahedral clusters may act to slow some of their neighboring particles to form regions of slow dynamics, which appear correlated with spatial patterns of dynamic heterogeneities [Fig. 4(a)], although this should be carefully checked.

\section{Summary}

We have demonstrated that studying the lifetimes of structural ordering within the Wahnstr\"{o}m  supercooled liquid can be used to directly identify the order that constitutes the slow regions of dynamic heterogeneities. 
The structures responsible, the icosahedra, combine to form non-space filling domains on cooling that become more stable as their size increases. 
Here, the crystal structure also has an icosahedral component, but icosahedra compatible with the crystal are rarely found in the supercooled liquid. Particles that neighbor the domains have suppressed mobility. These results imply that growing domains of icosahedra play an important in the microscopic origins of dynamic heterogeneities in this system. 
At the same time, the lack of a direct link in the length scales associated with structure and dynamics requires further 
study on the origin of slow dynamics including the mechanism of propagation of slowness from icosahedral clusters 
to their neighbors.

\section*{Acknowledgements}

We thank M. Leocmach for his kind help in bond orientational order analysis and its comparison with TCC analysis. 
A.M. is funded by EPSRC grant code EP/E501214/1. C.P.R. thanks the Royal Society for funding. 
H.T. acknowledges support from a grant-in-aid from the 
Ministry of Education, Culture, Sports, Science and Technology, Japan and the 
Aihara Project, the FIRST program from JSPS, initiated by CSTP. 
This work was carried out using the computational facilities of the Advanced Computing Research Centre, University of Bristol.

%\appendix{}

%\bibliography{alex_ref,alexJCPNotes}

\end{document}